\magnification=1200
\tenrm
\hsize=14truecm
\vsize=21.6truecm

\def\deg{^{\circ}}
\def\kmpersec{\,\rm\,km\,s^{-1}}
\def\etal{{\it et al. }}
\tolerance 1000

\hyphenpenalty 1000
\parindent 20pt
\parskip 5pt
\vskip 0.5truein
\centerline{\bf Galaxies Discovered Behind the Milky Way}
\centerline{\bf by the Dwingeloo Obscured Galaxies Survey}
\bigskip
\centerline{P. A. Henning}
\centerline{Institute for Astrophysics, University of New Mexico}
\centerline{800 Yale Blvd., NE, Albuquerque, NM 87131}
\medskip
\centerline{R. C. Kraan-Korteweg}
\centerline{Observatoire de Paris-Meudon, D.A.E.C.}
\centerline{92195 Meudon Cedex, France}
\centerline{and}
\centerline{Astronomy Department, University of Guanajuato}
\centerline{Apartado Postal 144, Guanajuato, GTO 36000, Mexico}
\medskip
\centerline{A. J. Rivers}
\centerline{Institute for Astrophysics, University of New Mexico}
\centerline{800 Yale Blvd., NE, Albuquerque, NM 87131}
\medskip
\centerline{A. J. Loan}
\centerline{Institute of Astronomy, University of Cambridge}
\centerline{Madingley Road, Cambridge CB3 0HA, UK}
\medskip
\centerline{O. Lahav}
\centerline{Institute of Astronomy, University of Cambridge}
\centerline{Madingley Road, Cambridge CB3 0HA, UK}
\centerline{and}
\centerline{Anglo Australian Observatory}
\centerline{P.O. Box 296, Epping, NSW 2121, Australia}
\medskip
\centerline{W. B. Burton}
\centerline{Leiden Observatory}
\centerline{Postbus 9513, NL-2300 RA Leiden, The Netherlands}
\vfill\eject
\noindent{Abstract}
\smallskip
Our Galaxy blocks a significant portion of the extragalactic sky from
view, hampering studies of large-scale structure.
This produces an incomplete knowledge of the distribution of galaxies,
and, assuming galaxies trace mass, of the gravity field.
Further, just one unrecognized, nearby massive galaxy could have large
influence over the Milky Way's motion with respect to the Cosmic
Microwave Background.
\par
Diligent surveys in the optical and infrared wavebands can find galaxies
through moderate Galactic gas and dust, but close to the Galactic Plane,
only radio surveys are effective. 
The entire northern Zone of Avoidance is being searched at 21 cm for
galaxies using the Dwingeloo 25-m telescope.
A shallow search for nearby, and/or 
massive galaxies has been completed, yielding
five objects.
Two of these galaxies were previously unknown, and although they are not likely
members of the Local Group, are part of the nearby Universe.
A deeper search continues, which will produce a flux-limited catalog of
hidden galaxies.
This portion of the survey is one-third complete, and
has detected about 40 objects to date.
Based on present understanding of the HI mass function, the complete survey
should uncover 50 - 100 galaxies.
\bigskip
\noindent{I. Introduction}
\smallskip
The dust and high star density of the Milky Way galaxy obscures
about 25\% of the optical extragalactic sky, and about 10\% in the
infrared,
producing a Zone of Avoidance
(ZOA).
The resulting incomplete coverage of surveys of external galaxies compromises
studies of large-scale structures.
The connectivity of superclusters and voids across the ZOA can be at present
only poorly studied.
Also, the all-sky distribution of mass which produces the observed dipole
in the Cosmic Microwave Background has not been fairly mapped, and, assuming
galaxies trace mass, filling in the missing information should produce
a fairer representation of the gravity field.
In particular, unrecognized nearby galaxies could have large influence,
as the nearest galaxies may generate a sizable fraction of the dipole moment
(Kraan-Korteweg 1989).
\par
Careful searches in the optical and infrared wavebands can narrow the
ZOA (see the many contributions in {\it Unveiling Large-Scale
Structures behind the Milky Way} 1994), but in the regions of highest
obscuration and infrared confusion near the Galactic Plane, only
radio surveys can find galaxies.
The 21-cm line of neutral hydrogen passes through the obscuration readily,
so galaxies with sufficient HI can be found through detection of their
21-cm emission.
Of course, this method will miss HI-poor, early-type galaxies.
The only ZOA which remains at 21 cm for HI-rich objects
comes from the inability to
discriminate extragalactic sources with redshifts near zero velocity
from Galactic HI.
Until recently, radio receivers were not sensitive enough for observers
to contemplate a full survey of the ZOA.
The first attempt to find hidden galaxies was a pilot survey with the
late 300-ft telescope, covering only $1.5\%$ of the ZOA visible from
Green Bank (Kerr \& Henning 1987; Henning 1992).
This survey proved the efficacy of the method, uncovering 19 previously
unknown objects, but the small beamsize of the 300-ft (and its subsequent
demise) made a full survey with that instrument unfeasible.
\par
To begin filling in the gaps in our knowledge of the distribution of
optically-hidden galaxies, a full survey of the northern ZOA is being conducted
with the 25-m radio telescope of the Netherlands Foundation for Research
in Astronomy, in Dwingeloo.
We report here on the results of a shallow, quick search 
(RMS noise per channel, $\sigma_{\rm ch}$ = 175 mJy)
for nearby,
massive galaxies, and on the status of a deeper search 
($\sigma_{\rm ch}$ = 40 mJy) 
for spirals
to a redshift of $4000\kmpersec$.
The two surveys together are referred to as the Dwingeloo Obscured Galaxies
Survey.
\medskip
\noindent{II.  Survey Strategy}
\medskip
\noindent{A.  The Shallow Search}
\smallskip
The first phase of the survey was conducted with the aim of finding
nearby and/or very massive galaxies.
Even a single nearby, unrecognized M31-type galaxy would have an
important gravitational effect on the Milky Way.
To uncover such a galaxy, the entire ZOA
accessible to the Dwingeloo telescope
was surveyed with 5-minute integration times within the redshift
range $V_{\rm LSR} = 0 - 4000\kmpersec$, i.e. approximately
15000 partially overlapping pointings,
with HPBW $0^{\circ}\!\!.6$
within 
$30^{\circ} \le\ l \le 220^{\circ}$;
$\vert b \vert \le 5^{\circ}\!\!.25$.
Negative velocities were not searched, as the Leiden/Dwingeloo
Galactic HI survey had recently covered the area, searching the range
$-450 \le V_{LSR} \le 400\kmpersec$ (Hartmann 1994, Hartmann \& Burton 1997), albeit with
higher RMS due to shorter integration time per pointing.
The obscured-galaxies survey utilizes the DAS-1000 channel digital autocorrelator
spectrometer developed by Bos (1989)
over 20 MHz bandwidth, producing a velocity resolution of
$4\kmpersec$, quite sufficient to distinguish narrow-linewidth
galaxies from monochromatic noise spikes.
(See Hartmann 1994, and Hartmann \& Burton 1997 for further details
of the telescope and equipment set-up, as used for both the
Galactic survey and the obscured-galaxies survey.)
The obscured-galaxies survey was conducted in total-power mode, with ON-OFF
pairs created from blind-survey pointings.
The exact survey strategy evolved over the course of the quick survey.
For the first portion, covering the upper third of the survey area,
($b > 1^{\circ}\!\!.4$), lines of sight were observed almost randomly,
using software which had been developed by Dap Hartmann for the Leiden/Dwingeloo Galactic
HI survey.
The order of pointings was chosen by an algorithm which determined
the length of time a position was above the horizon, preferentially
observing those positions which were visible for only short periods.
While this was fine for frequency-switched spectra, as was the case for
the Galactic survey, it produced non-optimal ON-OFF pairs for the obscured-galaxies survey,
as pointings observed close in time were not necessarily spatially close,
and vice versa.
For the bulk of the survey ($-5^{\circ}\!\!.25 \le b \le 1^{\circ}\!\!.05$),
observations were grouped into blocks of five consecutive pointings
at constant {\it b}, with {\it l} increased by $0^{\circ}\!\!.4$ increments.
From the sequence of pointings 1, 2, 3, 4, 5, ON-OFF pairs were created
as follows:  1-3, 2-4, 3-5, 4-1, 5-2.
The resulting spectra proved to be quite satisfactory, and it is this
strategy which also forms the basis of the deeper (one hour per pointing) survey
which is currently being carried out, and described below.
Using this method, a real galaxian signal will appear twice, once as a
positive signal, and again as a negative one, referenced against
two independent scans.
These blocks of five scans are organized into a honeycomb pattern
with $\Delta b = 0^{\circ}\!\!.35$ and adjacent rows offset by
$\Delta l = 0^{\circ}\!\!.2$.
Over most of the survey, the sensitivity obtained was 175 mJy
channel-to-channel variation.
The sensitivity to HI mass as a function of distance
can be derived
from the expression
$$M_{\rm HI} = 2.36 \times 10^5 r^2 \int S(V) {\rm d}V \, {\rm M_\odot},\eqno(1)$$
where {\it r} is the distance in Mpc and $\int S(V) {\rm d}V$ is the integral
over the line of the flux density in $\rm Jy\,\kmpersec$.
From past experience with HI surveys (e.g. Henning 1992, Kraan-Korteweg \etal 1997)
we note that a galaxy is detected
if its profile height satisfies
$$F_{\rm ob} \bar{S} > \nu_{\rm ch} \sigma_{\rm ch},\eqno(2)$$
where $\bar{S}$ is the mean flux density at 21 cm and
$\nu_{\rm ch}$ defines the threshold for detection (e.g.
$\nu_{\rm ch} = 2.5$).
$F_{\rm ob}$ is the `off-beam' fudge factor,
taking into account that a galaxy may lie away from the center of the
telescope beam in a blind search, and hence will have a reduced observed
flux.
For a Gaussian beam, $F_{\rm ob} = 0.7$.
The sensitivity of HI mass derived under these assumptions is shown in
Figure 1.
Note that other detection criteria are possible, e.g. in terms of the
{\it integrated} flux (Schneider 1996).
\par
The performance of the system was checked periodically by observing
NGC 3359 as a calibration source.
Of 14,725 planned pointings, 12647 have been observed, equivalent to
86\% of the 2000 deg$^2$ of the total survey area.
The latitude range $1^{\circ}\!\!.4 \le b \le 5^{\circ}\!\!.25$
was 72\% completed, and the range $-5^{\circ}\!\!.25 \le b \le 1^{\circ}\!\!.05$
was 94\% observed, the difference arising from our increased sophistication
of observing method for the lower portion.
Missed pointings were scattered evenly through the area,
save two significant gaps of $10^{\circ} \times 1^{\circ}$ in l, b (Fig. 2).
In addition, about 10\% of the scans were rendered useless by solar or
manmade interference.
A further 20\% were compromised over a narrow velocity range
($V_{LSR} = 1000 \pm 200\kmpersec$)
by fairly monochromatic interference.
The shape of this spurious signal was often irritatingly galaxian.
This significantly lowered sensitivity to galaxies at this redshift
since we were swamped by hundreds of false ``galaxies" at $1000\kmpersec$.
However, we could search these scans successfully over the rest
of the velocity band.
\par
Data from the telescope were translated into standard FITS format,
using software developed by Dap Hartmann, and the ON-OFF pairs created as above.
Blocks of five ON-OFF spectra were presented as postscript files for 
visual inspection.
When candidate signals were identified, adjacent pointings were
examined since it was possible that a galaxy could be detected in more
than one pointing.
For scans containing potential galaxian signals, the FITS data were loaded
into the Spectral Line Analysis Package (Staveley-Smith 1985)
for further analysis.
\medskip
\noindent{B.  The Deep Search}
\smallskip
After the shallow blind survey was completed, the telescope began
the deep phase of the survey, with the aim of producing a flux-limited
catalog.
The strategy was much the same, with blocks of five observations of
five minutes per pointing, but now these blocks are repeated twelve
times to reach an effective integration time of one hour per pointing,
and a sensitivity of 40 mJy.
This is not simply an improvement of $\sqrt {12}$ since we also switched to
a two-bit correlator sampling mode, further improving sensitivity.
The resulting sensitivity to HI mass is shown in Fig 3.
For example, for a linewidth of $100\kmpersec$ we are sensitive to galaxies 
with $5 \times 10^8 {\rm h}^{-2} {\rm M_\odot}$ of HI out to $10 {\rm h}^{-1}$ Mpc and
$5 \times 10^9 {\rm h}^{-2} {\rm M_\odot}$ of HI out to the edge of the survey's velocity
range, $40 {\rm h}^{-1}$ Mpc.
\par
The first area to be deeply surveyed was the region 
$125^{\circ} \le l \le 145^{\circ}$; $-5^{\circ}\!\!.25 \le b \le 5^{\circ}\!\!.25$,
encompassing the crossing of the Supergalactic plane and the Galactic
plane.
This has been completed, and now the telescope is observing the rest of
the northern ZOA, starting at $b = -5^{\circ}\!\!.25$ and observing strips of increasing
latitude.
At the time of writing (June 1997) the deep survey had been completed
from $b = -5^{\circ}\!\!.25$ to $b = -2^{\circ}\!\!.1$.
\bigskip
\noindent{III.  Survey Results}
\smallskip
\noindent{A.  The Shallow Search}
\par
The shallow survey detected five galaxies, three of which were known
previously.
The sources are listed in Table 1, with HI parameters as observed 
during the shallow search period of the Dwingeloo survey. 
The positions of each pointing where the galaxies were detected are
given, along with the strength of the signal at each location.
For the position where the signal is at its strongest, the
basic parameters of the line are measured.
The extinction was derived from the Galactic
HI column density (Weaver \& Williams 1973) assuming the gas-to-dust 
ratio derived by
Burstein and Heiles (1982).
The profiles, shown in Figure 4a - e, are constructed from the 
sum of the scans where each galaxy was detected.
\par
Two of the previously known galaxies were originally discovered in the
infrared (Maffei 2:  Maffei 1968;  IRAS 05596+1451:  identified as a
galaxy by Lu \etal 1990).
Both have faint optical counterparts visible on the Palomar Sky Survey
prints, and are heavily obscured spirals.
The association of the Dwingeloo-detected objects with these cataloged galaxies
rests not only on the angular position, but also on the correspondence
with previous redshift measurements (Maffei 2:  e.g. Bottinelli \etal 1971,
Bottinelli \& Gougenheim 1976, Lewis \& Davies 1973, Love 1972,
Shostak \& Weliachew 1971, and others;
IRAS 05596+1451:  Lu \etal 1990.)
In the case of Maffei 2, previous 21-cm observations show that the
profile extends to negative velocities.
We do not recover all of its emission, since our search range covered
only positive velocities.
The other source known previously, [H92] 34, was discovered through a
pilot blind 21-cm survey of a small portion of the ZOA made with the
NRAO 300-ft telescope (Henning 1992).
HI synthesis observations conducted with the VLA confirmed the detection,
and indicated a mass and a profile shape consistent with the source being a
normal, but totally obscured, spiral.
\par
The other two sources detected are previously unknown, hidden galaxies.
The galaxy Dwingeloo 1, which we catalog here also as Dw138.5-0.1, 
has been the subject of much 
follow-up observation (Kraan-Korteweg \etal 1994, Loan \etal 1996,
Burton \etal 1996, Kuno \etal 1996, Li \etal 1996, Tilanus \& Burton 1997).
To summarize, it is a massive barred spiral,
with rotation velocity of $130\kmpersec$ compared to $220\kmpersec$ for the Milky Way,
implying a dynamical mass within a radius where the rotation curve
is flat of
roughly one-third the
mass of the Milky Way. 
The exact mass is inaccurately known
primarily due to its uncertain distance ($\approx\ 3 \,{\rm Mpc}$).
Its angular location and redshift place it within the IC342/Maffei
group of galaxies.
It has a dwarf companion, Dwingeloo 2, which was discovered through 
sensitive follow-up HI synthesis observations with the Westerbork Synthesis Radio
Telescope (Burton \etal 1996).
The identification of this small galaxy as a companion to Dwingeloo 1 rests on
its angular and kinematic proximity.
Also, Dwingeloo 2 shows signs of distortion, possibly due to
interaction with its more massive companion.
\par
The other galaxy discovered through the shallow search, Dw095.0+1.0
is also likely to be nearby, with a recession velocity 
$V_{\rm LSR} = 159\kmpersec$, or $V_{\rm G} = 378\kmpersec$ 
[where $V_{\rm G} = V_{\rm LSR} + 220\cos(b)\sin(l)$].
However, this source cannot be very massive.
Naively assuming a distance of 3 Mpc, the HI mass would be about
$8 \times 10^7 {\rm h}^{-2} {\rm M_\odot}$, bearing in mind that it may not be
at the center of the beam, and the true flux density may be somewhat
higher.
(Planned follow-up HI synthesis observations will pin down its location
and flux density.)
Its narrow profile (FWHM = $87\kmpersec$) supports the idea
that it is a hidden, nearby dwarf.
There is no optical counterpart visible, due to high obscuration.
Likewise, there is no obvious corresponding extragalactic IRAS source,
although this is not surprising as IRAS was not very sensitive to
dwarfs (Bothun \etal 1989; Sauvage \etal 1990), and confusion due to
Galactic IR sources often masks low-latitude galaxies found in HI
(Henning 1995).
Its angular position and recession velocity indicate that it is not
likely a member of the Local Group.
Plotting recession velocity versus distance from the solar apex as
determined by Sandage (1986), it lies well outside the range of
known and even suspected Local Group galaxies (see Fig 1 of
van den Bergh 1994).
It is also not likely a member of the IC342/Maffei/Dw1 group, since it lies
$\approx 40\deg$ from that group's center (Krismer \etal 1995).
However, it is close in velocity to NGC 6946 at $(l,\,b,\,V_{\rm hel}\, = 
95^{\circ}\!\!.7,\,11^{\circ}\!\!.7,\,49\kmpersec$; the velocity quoted is the average of
measurements by Dean \& Davies 1975, 
Gordon \etal 1968, Rogstad \& Shostak 1972, Rots 1980, Tacconi \& Young 1986,
Tifft \& Cocke 1988.)
Since truly isolated galaxies are rare, this raises the possibility
that these galaxies may be part of a previously unidentified nearby
group.
Such a group should be uncovered by the Dwingeloo Deep Search when it
reaches this latitude range (except for low velocity members whose
HI signals may be lost in Galactic HI.)
If this hypothetical group exists, it is offset from the Supergalactic
Plane by about $40^\circ$, considerably more than any other known group
in the local Universe.
\par
With the completion of the Shallow Search of the entire northern
ZOA for nearby, massive HI galaxies,
we can say with some confidence that the most massive, nearby, previously
unidentified galaxy detectable in HI in this search area is Dwingeloo 1.
However, since the survey covered only about 80\% of the
originally planned area, it is possible galaxies could have
gone undetected.
From an HI mass function, we estimate four galaxies should lie
in the full survey region (Section IV) so $(0.2) \times (4) = 1$
galaxy may lie in the unsurveyed area.
The chances of missing a large, nearby, and therefore, extended
galaxy are low, because such an object would likely be detectable
in several adjacent pointings, all of which would have to be missed
for the galaxy to escape detection.
Thus, it is fairly unlikely that there is another
previously 
unidentified, massive galaxy whose gravitational influence impacts
Local Group peculiar motion or internal dynamics in an important
way in the area of the survey.
\par
Finally, we consider the possibility of a large, unrecognized galaxy
lying outside the range of our survey.
From the distribution of Galactic HI (data from Weaver \& Williams 1973, map
presented by Bloemen 1983) and the conversion from HI column density to
extinction A$_{\rm B}$, assuming a constant Galactic gas-to-dust ratio
(Burstein \& Heiles 1978), we find the extinction falls to roughly 2.5 magnitudes
at $\vert b \vert = 5^{\circ}\!\!.25$, with some patches exceeding 4 magnitudes
of obscuration.
An A$_{\rm B}$ of 2.5 magnitudes corresponds to a diminution in angular size
of a factor of three (Cameron 1990).
A galaxy near enough and large enough to have significant gravitational influence,
an M31-like galaxy, would still be recognizable through this amount of obscuration.
Of course, we are truly blind to galaxies in the southern ZOA, longitude
range from l = $220^{\circ}$ to $30^{\circ}$, which lies below the declination
limit of the Dwingeloo telescope.
Within the same latitude range as the northern search, this corresponds to
$1800\,{\rm deg}^2$, comparable to the $2000\,{\rm deg}^2$ we surveyed in the north.
Thus, about four galaxies would be detected if this southern area were
surveyed with the search parameters used in this survey, as discussed in
Section IV.
A more sensitive 21-cm survey being conducted with the Parkes 64-m
telescope will cover this southern gap (Staveley-Smith 1996).
The remaining blind spot of any 21-cm survey for galaxies in the ZOA is the slice
in velocity space contaminated by Galactic HI, within roughly $\pm 200\kmpersec$,
with contamination level strongly dependent on l and b.
\par
To double check that the Shallow Search did indeed recover the nearby known
galaxies it should have, a search for all cataloged galaxies within our
angular and velocity restrictions was conducted using the Lyon-Meudon
Extragalactic Database (LEDA), and the NASA/IPAC Extragalactic Database
(NED).
This produced a list of five galaxies from other surveys
with systemic velocities
between $0 - 1000\kmpersec$.
Of the five, we did detect two (Maffei 2, Dw136.5-0.3, and IRAS05596+1451, Dw194.1-3.8).
The other three were not recovered for well understood reasons:
Maffei 1, an elliptical, which would not be detected in an HI survey;  
[H92] 11, a galaxy discovered through
an earlier HI survey (Henning 1992); and MB 1,
an HI-deficient, late type spiral in the IC 342/Maffei/Dw1
group (McCall \etal 1995; Huchtmeier \& van Driel 1996).
The flux densities of the latter two galaxies are
well below the sensitivity limit of the Shallow
Search, but they been observed and detected by the continuing Deep Search,
described below.
\medskip
\noindent{B.  The Deep Search}
\smallskip
Here we present a progress report on the Deep Search phase of the Dwingeloo
Obscured Galaxies Survey.
We have detected 40 galaxies over the 790 ${\rm deg}^2$ surveyed
to date.
Half of these galaxies lie in the region of the
Galactic Plane/Supergalactic Plane crossing.
The locations of the galaxies are shown in Fig. 5, along with the
positions of previously cataloged objects in the search region and
surrounding higher latitudes.
A few sources await confirmation with the Dwingeloo telescope.
A list of objects' positions and 21-cm flux densities will be published
after HI synthesis follow-up observations are completed.
With these data, we will investigate the properties of HI-selected
galaxies in different environments (e.g. Local Void vs. Local Supercluster)
and the observed HI mass function.
\bigskip
\noindent{IV.  Detection Predictions from an HI Mass Function}
\par
Here we attempt to predict the number of HI galaxies expected to be
found in a 21-cm survey, observed with the parameters being used for
the Dwingeloo work.
The prediction attempts to mimic the detection of galaxies by eye.
\par 
A prediction of the number of HI galaxies expected to be detected by a
survey requires some knowledge of the HI mass function, $\phi (M_{\rm HI})$.
The complete HI mass function will be a bivariate function of the HI 
mass and the velocity width, $\phi (M_{\rm HI}, \Delta V)$.  
However, the
present data sets are not sufficient to estimate the bivariate function,
and (following Rao \& Briggs 1993 and Schneider 1996) the HI mass
function is assumed to be approximated by a Schechter (1976) function for
every $\Delta V$,
$$\phi(M_{\rm HI}) {\rm d}M_{\rm HI} = \phi_* \left({M_{\rm HI}}\over{M_*}\right)^{-\alpha}
{\rm exp}\left(-{M_{\rm HI}}\over{M_*}\right)d\left({M_{\rm HI}}\over{M_*}\right),\eqno(3)$$
where $\alpha$ is the low-mass-end slope, $M_*$ is the characteristic HI 
mass, and $\phi_*$ is a normalization parameter.  
After Zwaan et al.\ 
(1997), we adopt $\alpha=1.25$, $ M_* = 0.4 \times 10^{10} {\rm M_{\odot}}$, 
$\phi_* = 0.013 $ galaxies per ${\rm (Mpc/h)}^3$.
\par
The derivation of the HI mass function by Henning (1995),
Rao \& Briggs (1993), Zwaan et al.\ (1997), and others uses the
$1/V_{max}$ method.  
This method makes the assumption that the galaxy
population is distributed in a homogeneous manner.  
However, galaxies are
clumped together in groups and clusters, 
hence this method produces a biased statistic.  
By using a clustering-independent parametric
maximum-likelihood technique, such as the STY technique (after Sandage \etal 1979),
the HI mass function may be more fairly derived.
This method has been used recently for an optically-selected sample
(Solanes \etal 1996.)
Work is underway to construct HI mass functions in this manner for various
HI-selected samples (Lahav \etal in prep.)
\bigskip
\noindent{A.  The Expected Number of Galaxies}
\par
For a uniform distribution, the number of galaxies expected to be detected
in a survey out to a distance $R_{max}$ with solid angle
$\Omega_{s}$ can be calculated by integrating over the mass
function:
$$N(< R_{max}) = \Omega_{s} \; \int_0^{R_{max}} {\rm d}r r^2 \;
  \int_{x_{min}}^\infty \phi(x) dx,\eqno(4)$$
where $x= M_{\rm HI} / M_*$, and the minimum HI mass to be observed
(from Equation 1) is
$$M_{\rm HI, min} = x_{min} M_* = 2.36 \times 10^5 
  \left({\Delta V \nu_{\rm ch} \sigma_{\rm ch}}\over{F_{\rm ob}}\right)
  \; r^2 \; M_{\odot}.\eqno(5)$$
As before, $\sigma_{\rm ch}$ is the RMS noise per channel,
$\nu_{\rm ch}$ defines the threshold for detection (e.g.
$\nu_{\rm ch} = 2.5$),
and $F_{\rm ob}$ is the factor
which takes into account that a galaxy may lie away from the center of the
telescope beam in a blind search, and hence will have a reduced observed
flux.
\par
Assuming that the mass function is a Schechter function, then
$$\int_{x_{min}}^\infty \phi(x) {\rm d}x = 
  \phi_* \; \Gamma \left[1-\alpha, x_{min}(r)\right]\;,\eqno(6)$$
where $\Gamma$ is the incomplete Gamma function.  Thus, the
result depends on the combination of parameters $(\Delta V \nu_{\rm ch}
\sigma_{\rm ch}) / (F_{\rm ob} M_*)$.
\par 
If the survey extends to infinity ($R_{max} = \infty$) then the
integral in Equation 4 can be solved analytically:
$$N(< \infty) = 2.9 \times 10^{-9} \; \Omega_{s} \; \phi_* 
\Gamma\left( {{5}\over{2}} - \alpha \right)
\left({M_*}\over{M_{\odot}}\right)^{{3}\over{2}} 
\left({\Delta V \nu_{\rm ch}\sigma_{\rm ch}}\over{F_{\rm ob}}\right)^{{-3}\over{2}},\eqno(7)$$
where $\Gamma$ is the complete Gamma function; if $\alpha=1.25$ then
$\Gamma\left( (5/2) - \alpha\right) = 0.906$.  
This calculation extends to
infinity, and thus the predicted number of galaxies will be higher than the
number expected within a survey with a finite maximum depth.  
However, it
does produce a fiducial number of galaxies.  
Once the Dwingeloo Deep Search is complete, this
will provide an estimate for the number of galaxies that would have been
detected if they did not lie beyond $R_{max}$.
\bigskip
\noindent{B.  Results and Predictions} 
\par
The appropriate parameters for the survey are: $R_{max} =
4000\kmpersec$, $\sigma_{\rm ch} = 175$ mJy (for the Shallow Survey),
$F_{\rm ob} = 1$, $\nu_{\rm ch}=2.5$.
Assuming a fiducial value of $\Delta V =
185\kmpersec$ (the mean value found by Henning (1995) for an HI-selected
sample) and the HI mass function
parameters from Zwaan et al.\ (1997), then $N = 8$ galaxies should be
detected in the $2000\,{\rm deg}^2$ survey area ($\Omega_{s}=0.6$ str).  
The number of expected galaxies is the same if $R_{max} =
\infty$: galaxies cannot be detected beyond the survey limits with an
integration time of only $5\,{\rm min}$.  
Remembering the off-beam factor, if
$F_{\rm ob} = 0.7$ then the predicted number in the Shallow Survey
is $N=4$ galaxies.
The observed number is five galaxies, very close to this prediction.
\par
For the Deep Search portion of the survey ($\sigma_{\rm ch} = 40$ mJy) the
corresponding predictions are $N=62$ galaxies for $F_{\rm ob}=1.0$, and
$N=39$ galaxies for $F_{\rm ob}=0.7$.  
However, these calculations take
no account of density enhancements within the survey area.  The area chosen
for the survey includes regions that are known to be over-dense.  
For example,
the Supergalactic Plane is 10--40~per~cent denser than the uniform
background (Lahav et al.\ in prep).  
Thus, the full Dwingeloo Deep Survey should be expected to
find 50--100 galaxies.
This is quite in line with the detection of 40 galaxies at this stage of
survey completion, $790\,{\rm deg}^2$ observed of the $2000\,{\rm deg}^2$
total area to be done.
\bigskip
\noindent{V.  Summary and Future Directions}
\par
A shallow search for nearby, and/or massive galaxies detectable at 21-cm
has been completed over the northern ZOA, with the detection of five
galaxies.  
Two of these were previously unknown, and while their distances are
not as yet very accurately determined, they are part of the local Universe.
A more sensitive search continues, which should uncover 50 - 100 galaxies
to the edge of the survey range, $R_{max} = 4000\kmpersec$.
\par
The southern ZOA is currently being surveyed at 21-cm with the new multibeam
receiver system on the Parkes radiotelescope (Staveley-Smith 1997).
Extrapolating from observed HI mass functions based on only dozens of
galaxies, as in Section IV, this survey should uncover thousands of galaxies,
due to its excellent sensitivity and deep velocity coverage, $R_{max} = 12,200\kmpersec$.
For the present Dwingeloo Survey, the galaxies are uncovered by visual inspection
of the spectra. 
An alternative approach for handling future HI
surveys, like the Parkes multibeam survey,
is to attempt to devise an automated galaxy 
detection algorithm, e.g. Loan (1997).
\bigskip
\noindent{Acknowledgments}
\par
We are grateful to D. Hartmann for the development of special-purpose
software and for running a variety of tests which established the
feasibility of the Dwingeloo Obscured Galaxies Survey; to G. Hau
and H. Ferguson for their early work in getting the survey running
routinely, to A. Foley and D. Moorrees for their supervision of the
day-to-day telescope operations, and to D. Lynden-Bell for helpful
discussions.
\par
The research has made use of the NASA/IPAC Extragalactic Database (NED)
which is operated by the Jet Propulsion Laboratory, California
Institute of Technology, under contract with the National Aeronautics
and Space Administration.
We have also made use of the Lyon-Meudon Extragalactic Database (LEDA)
supplied by the LEDA team at the CRAL-Observatoire de Lyon (France.)
The Dwingeloo 25-m radiotelescope is supported by the Netherlands
Foundation for Scientific Research (NWO).
\par
The research of PH is supported by NSF Faculty Early Career
Development (CAREER) Program award AST-9502268.
The research of RCKK has been supported by an EC grant.
\bigskip
\noindent{References}
\smallskip\noindent
Bloemen, J.B.G.M. 1983, in {\it Surveys of the Southern Galaxy}, 
ApScS Library, vol. 105, eds.
W.B. Burton \& F.P. Israel, Reidel: Dordrecht
\smallskip\noindent
Bos, A. 1989, The 1024 Channel Spectrometer for the Dwingeloo Telescope,
Internal Technical Report, 188, NFRA
\smallskip\noindent
Bothun, G.D., Lonsdale, C.J., \& Rice, W. 1989, ApJ, 341, 129
\smallskip\noindent
Bottinelli, L., Chamaraux, P., Gerard, E., Gouguenheim, L, Heidmann,
J., Kazes, I., \& Lauque, R. 1971, A\&A, 12, 264
\smallskip\noindent
Bottinelli, L., \& Gouguenheim, L. 1976, A\&A, 47, 381
\smallskip\noindent
Burstein, D., \& Heiles, C. 1978, ApJ, 225, 40
\smallskip\noindent
Burstein, D., \& Heiles, C. 1982, AJ, 87, 1165
\smallskip\noindent
Burton, W.B., Verheijen, M.A.W., Kraan-Korteweg, R.C., \& Henning,
P.A. 1996, A\&A., 309, 687
\smallskip
\noindent
Cameron, L.M. 1990, A\&A, 233, 16
\smallskip\noindent
de Vaucouleurs, G., de Vaucouleurs, A., \& Corwin, H.G. 1976, Second
Reference Catalogue of Bright Galaxies (Austin:  Univ. of Texas Press)
\smallskip\noindent
Dean, J.F., \& Davies, R.D. 1975, MNRAS, 170, 503
\smallskip\noindent
Gordon, K.J., Remage, N.H., \& Roberts, M.S. 1968, ApJ, 154, 845
\smallskip\noindent
Hartmann, D. 1994, Ph.D. Thesis, R.U. Leiden
\smallskip\noindent
Hartmann, D., \& Burton, W.B. 1997, Atlas of Galactic Neutral Hydrogen,
Cambridge University Press
\smallskip\noindent
Henning, P.A. 1992, ApJS, 78, 365
\smallskip\noindent
Henning, P.A. 1995, ApJ, 450, 578
\smallskip\noindent
Huchtmeier, W.K., \& van Driel, W. 1996, A\&A, 305, 25
\smallskip\noindent
Kerr, F.J., \& Henning, P.A. 1987, ApJ, 320, L99
\smallskip\noindent
Kraan-Korteweg, R.C.  1989, in {\it Rev. in Modern Astron.} 2,
ed. G. Klare, Springer:  Berlin
\smallskip\noindent
Kraan-Korteweg, R.C., Woudt, P.A., \& Henning, P.A. 1997, PASA, 14, 15
\smallskip\noindent
Kraan-Korteweg, R.C., Loan, A.J., Burton, W.B., Lahav, O.,
Ferguson, H.C., Henning, P.A., \& Lynden-Bell, D. 1994,
Nature, 372, 77
\smallskip\noindent
Krismer, M., Tully, R.B., \& Gioia, I.M. 1995, AJ, 110, 1584
\smallskip\noindent
Kuno, N., Vila-Vilaro, B., \& Nishiyama, K. 1996, PASJ, 48, 19
\smallskip\noindent
Lewis, B.M., \& Davies, R.D. 1973, MNRAS, 165, 213
\smallskip\noindent
Li, J.G., Zhao, J.H., Ho, P.T.P., \& Sage, L.J. 1996, A\&A, 307, 424
\smallskip\noindent
Loan, A.J. 1997, Ph.D. Thesis, University of Cambridge
\smallskip\noindent
Loan, A.J., Maddox, S.J., Lahav, O., Balcells, M., Kraan-Korteweg,
R.C., Assendorp, R., Almoznino, E., Brosch, N., Goldberg, E.,
\& Ofek, E.O. 1996, MNRAS, 280, 537
\smallskip\noindent
Love, R. 1972, Nature, 235, 53
\smallskip\noindent
Lu, N.Y., Dow, M.W., Houck, J.R., Salpeter, E.E., \& Lewis, B.M.
1990, ApJ, 357, 388
\smallskip\noindent
McCall, M.L., Buta, R.J., \& Huchtmeier, W.K. 1995, IAU Circular
No. 6159
\smallskip\noindent
Maffei, P. 1968, PASP, 80, 618
\smallskip\noindent
Nilson, P. 1973, Uppsala General Catalogue of Galaxies, Uppsala
Astron. Obs. Ann., 6
\smallskip\noindent
Rao, S., \& Briggs, F.H. 1993, ApJ, 419, 515
\smallskip\noindent
Rogstad, D.H., \& Shostak, G.S. 1972, ApJ, 176, 315
\smallskip\noindent
Rots, A.H. 1980, A\&AS, 41, 189
\smallskip\noindent
Sandage, A. 1986, ApJ, 307, 1
\smallskip\noindent
Sandage, A., Tammann, G.A., \& Yahil, A. 1979, ApJ, 232, 352
\smallskip\noindent
Sauvage, M., Thuan, T.X., \& Vigroux, L. 1990, A\&A, 237, 296 
\smallskip\noindent
Schechter, P. 1976, ApJ. 203, 297
\smallskip\noindent
Schneider, S.E. 1996, The Minnesota Lectures on Extragalactic
Neutral Hydrogen, ASP Conf. Ser. 106, ed. E.D. Skillman, p. 323
\smallskip\noindent
Shostak, G.S., \& Weliachew, L 1971, ApJ, 169, L71
\smallskip\noindent
Solanes, J. M., Giovanelli, R., \& Haynes, M. P. 1996, ApJ,
461, 609
\smallskip\noindent
Staveley-Smith, L. 1985, Ph.D. Thesis, Univ. of Manchester 
\smallskip\noindent
Staveley-Smith, L. 1997, PASA, 14, 111
\smallskip\noindent
Tacconi, L.J., \& Young, J.S. 1986, ApJ, 308, 600
\smallskip\noindent
Tifft, W.G., \& Cocke, W.J. 1988, ApJS, 67, 1
\smallskip\noindent
Tilanus, R.P.J., \& Burton, W.B. 1997 A\&A, in press
\smallskip\noindent
{\it Unveiling Large-Scale Structures behind the Milky Way},
eds. C. Balkowski and R.C. Kraan-Korteweg, 1994, ASP Conf. Ser. 67
\smallskip\noindent
van den Bergh, S. 1994, AJ, 107, 1328
\smallskip\noindent
Weaver, H.F., \& Williams, D.R.W. 1973, A\&AS, 8, 1
\smallskip\noindent
Zwaan, M., Briggs, F., \& Sprayberry, D. 1997, PASA, 14, 126
\vfill\eject

\def\firstline {\vskip 5pt
\hrule
\vskip 2pt
\hrule
\vskip 5pt}
\def\secondline {\noalign{
\vskip 5pt
\hrule
\vskip 5pt}}
$$\vbox{
\hbox {Table 1.  Galaxies Detected through the Dwingeloo Shallow Search}
\firstline
\halign {#\hfil &\hfil#\hfil&\hfil#\hfil&\hfil#\hfil&
\hfil#\hfil&\hfil#\hfil&\hfil#\hfil&\hfil#\hfil&\hfil#\hfil\cr
Galaxy\quad&
\quad l \quad&\quad b \quad & 
\quad V$_{\rm LSR}$ \quad&\quad $\Delta$V$_{50}$ &\quad Peak \quad 
&\quad HI Flux \quad &
\quad A$_V$ \quad & \quad Opt? \quad \cr
& (deg) & (deg) & ($\kmpersec$) & ($\kmpersec$) & (Jy) & (Jy $\kmpersec$) 
& (mag) \cr
\secondline
Dw095.0+1.0 &  95.0  & 1.1 & 159 &  87 & 0.60 & 38 & 6.5 & N\cr 
              &  95.4  & 1.1 &     &     & 0.45 &    &     &  \cr
              &        &     &     &     &      &    &     &  \cr
[H92] 34 &  95.0  & 1.8 & 1046 & 93 & 0.96 & 77 & 6.5 & N\cr
 = Dw095.1+1.6   &  95.2  & 1.4 &     &     & 0.69 &    &     & \cr
	      &  95.4  & 1.8 &     &     & 0.59 &    &     & \cr
              &        &     &     &     &      &    &     &  \cr
Maffei 2 & 136.6  & -0.4 & 98 & 84 & 0.88 & 53 & 4.3 & Y\cr
 = Dw136.5-0.3   & 136.4  & -0.7 &    &    & 0.37 &    &     & \cr
              &        &     &     &     &      &    &     &  \cr
Dw1 & 138.6  & -0.4 & 110 & 184 & 1.31 & 152 & 5.3 & Y\cr
 = Dw138.5-0.1        & 138.8  & 0.0 &      &     & 1.18 &     &    & \cr
              & 138.2  & -0.4 &     &     & 1.08: &    &    & \cr
	      & 138.4  & 0.0 &      &     & 1.00  &    &    & \cr 
              &        &     &     &     &      &    &     &  \cr
IRAS 05596+1451 & 194.2  & -3.9 & 745 & 231 & 0.87  & 131 & 2.7 & Y\cr
 = Dw194.1-3.8 & 194.0 & -3.5   &    &  & 0.58 & & & \cr
		   & 194.4 & -3.5    &    &  & 0.56 & & & \cr 
		   & 194.6 & -3.9   &    &  & 0.46 & & &\cr}
\vskip 5pt
\hrule}$$
\noindent
Note:  For all galaxies except Dw095.0+1.0, optical or IR identification,
or HI synthesis observations have provided accurate positions, which are
reflected in the galaxies' names.
For Dw095.0+1.0, the name is based on the search point where it
shows the strongest signal, which is not necessarily the galaxy's true
position.
\smallskip
\noindent
The uncertainty in peak flux for the third position quoted for Dw138.5-0.1
is due to interference.
\vfill\eject

\noindent{Figure Captions}
\par\noindent
Figure 1.  Sensitivity to HI mass as a function of distance for three
representative linewidths for the Shallow Search, with integration
time of 5 minutes per pointing.
The channel-to-channel RMS variation achieved is 175 mJy,
the threshold for detection is taken to be $2.5$ times the
RMS noise over the profile.
Account is taken of the random placement of a galaxy in the
Gaussian beam, which lowers the observed flux to an average
of 70\% of the beam-center value.
\par\noindent
Figure 2.  Coverage map of the Shallow Search.
Over the latitude range 
$1^{\circ}\!\!.4 \le b \le 5^{\circ}\!\!.25$, the survey was 72\% complete.
From $-5^{\circ}\!\!.25 \le b \le 1^{\circ}\!\!.05$, 94\% of the
planned pointings were observed.
About 10\% of scans were badly disturbed by interference.
\par\noindent
Figure 3.  Sensitivity to HI mass as a function of distance for three
representative linewidths for the Deep Search, with integration time of
1 hour per pointing.
The channel-to-channel RMS variation achieved is 40 mJy,
the threshold for detection is again taken to be $2.5$ times the
RMS noise over the profile.
As in Fig. 1, account is taken of the random placement of a galaxy in the
Gaussian beam.
\par\noindent
Figure 4.  HI profiles of the five galaxies detected by the Shallow
Search.
The profiles show the sum of emission for all pointings where 21-cm
signal is detected for each galaxy.
\par\noindent
Figure 5.  Distribution of previously-cataloged galaxies with redshifts
measured to be within
$4000\kmpersec$ are shown as crosses, and the galaxies detected
in Dwingeloo are shown as squares (Shallow Survey) and triangles (Deep Survey).
The previously-cataloged galaxies  
mostly come from optical compilations [235 from the Uppsala General Catalog
(Nilson 1973) 
and 62 from the Second Reference Catalogue of Bright Galaxies (de
Vaucouleurs \etal 1976)].
The rest are from smaller optical surveys, or from IRAS.
The dotted lines indicate the area of the sky surveyed to date by
the Deep Survey. 
The Supergalactic Plane runs roughly vertically
through the Galactic Plane, crossing in the longitude range $130^{\circ} - 145^{\circ}$.
Two Shallow Survey detections lie in the IC 342/Maffei/Dw1 group at
$l \approx 140^{\circ}, b \approx 0^{\circ}$.
The Local Void is evident as a general underdensity of galaxies between
$l \approx 30^{\circ} - 75^{\circ}$.
Several galaxies detected by the Deep Survey lie in this region.
\vfill\eject
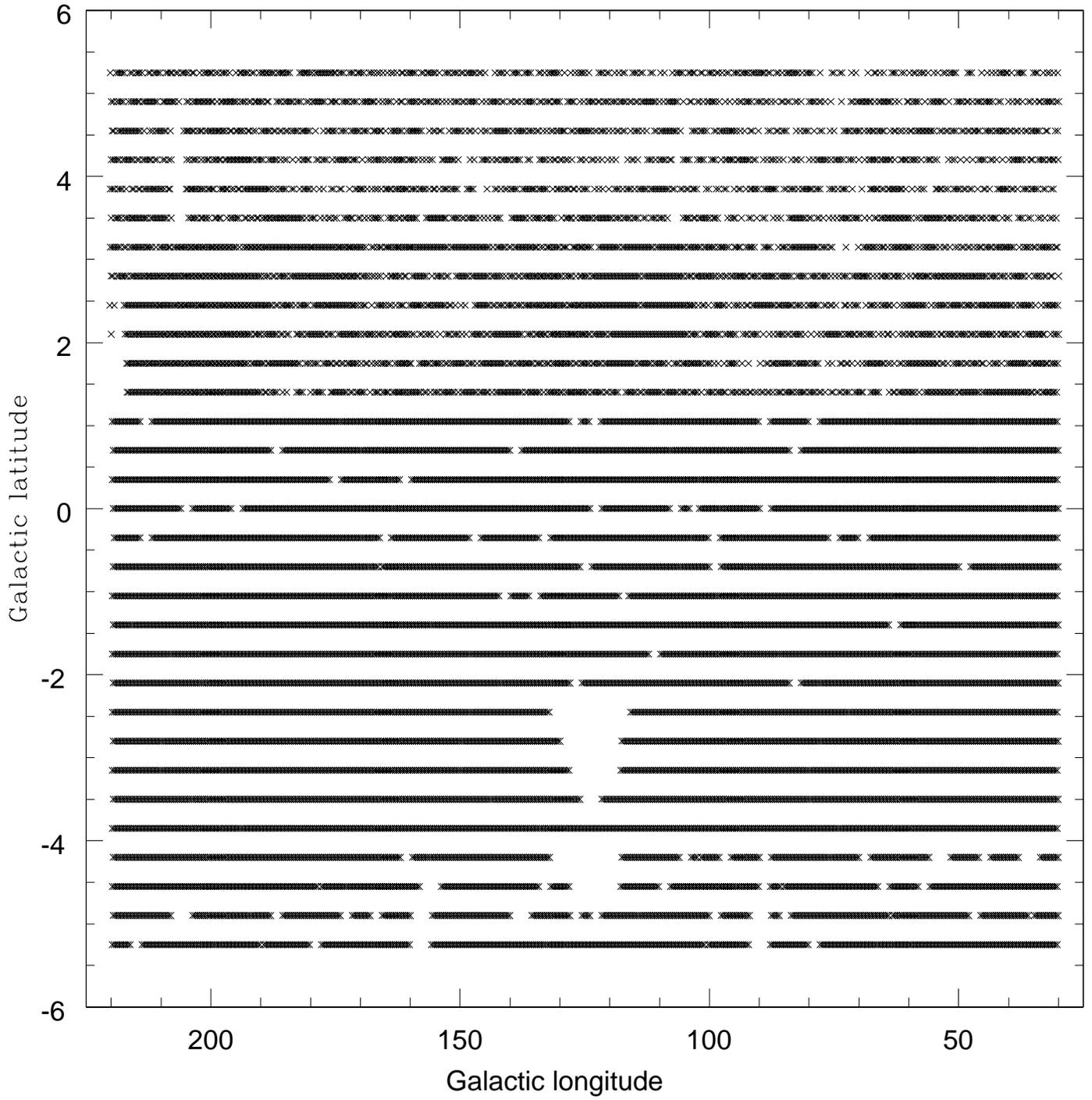
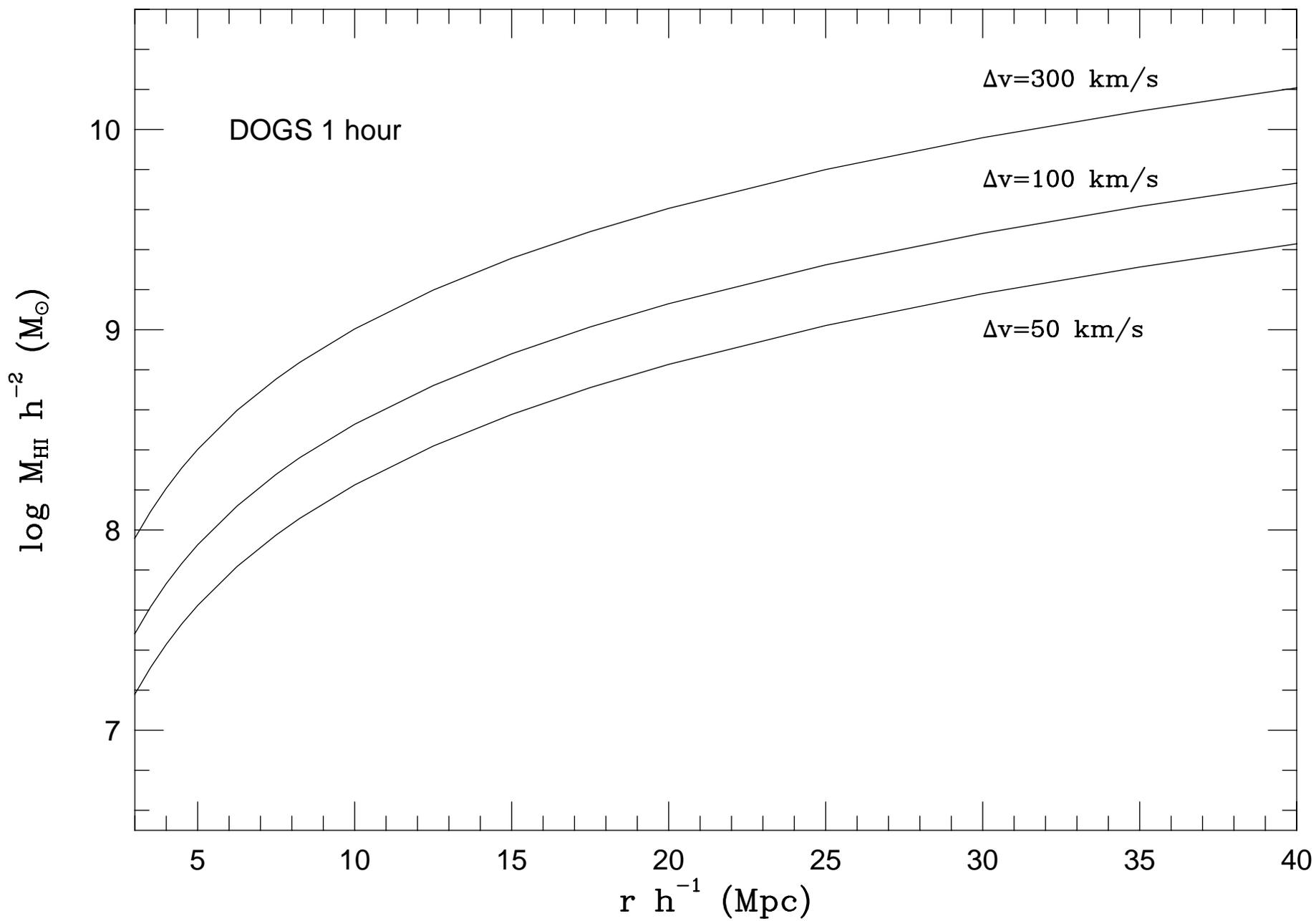
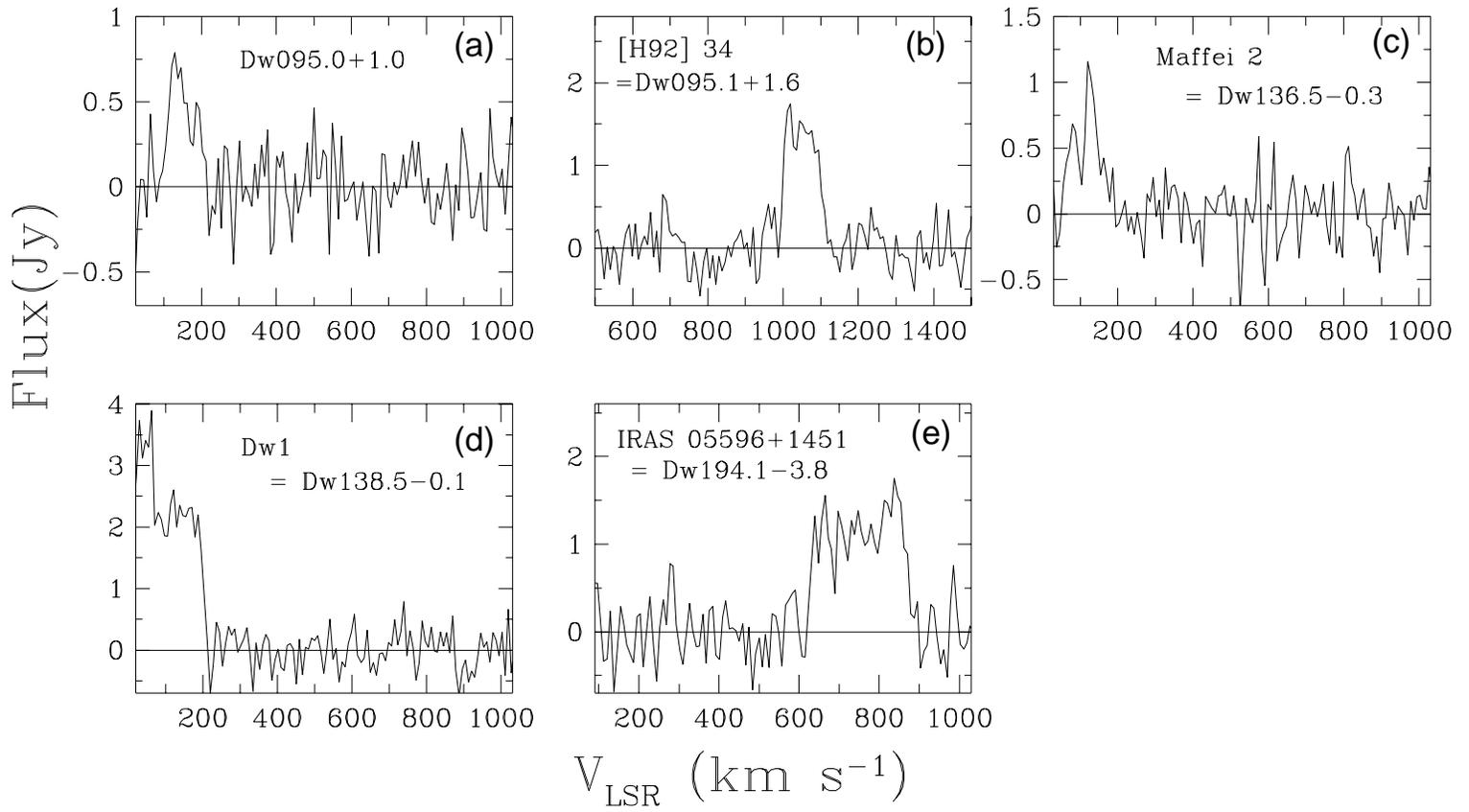
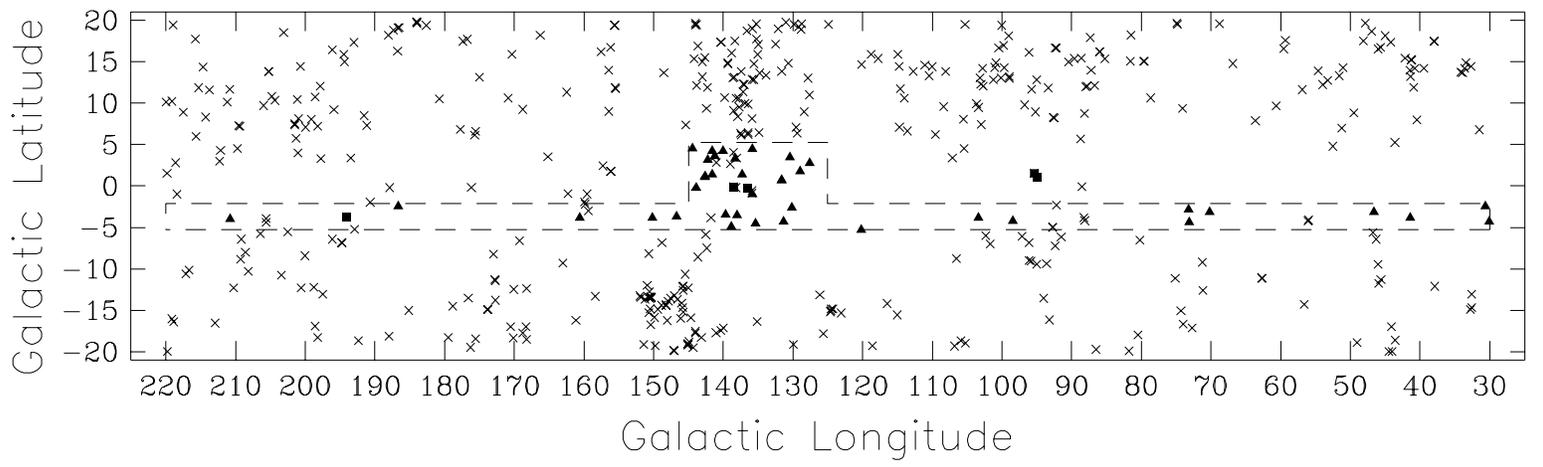
\end